\documentclass[fleqn,usenatbib]{mnras}
\usepackage[T1]{fontenc}

\DeclareRobustCommand{\VAN}[3]{#2}
\let\VANthebibliography\thebibliography
\def\thebibliography{\DeclareRobustCommand{\VAN}[3]{##3}\VANthebibliography}

\usepackage{graphicx}	% Including figure files
\usepackage{amsmath}	% Advanced maths commands
\newcommand{\Msun}{\mbox{M$_{\odot}$}}

\newcommand{\lppr}{\stackrel{<}{\scriptstyle \sim}}
\newcommand{\lappr}{\raisebox{-0.4ex}{$\lppr$}}
\newcommand{\gppr}{\stackrel{>}{\scriptstyle \sim}}
\newcommand{\gappr}{\raisebox{-0.4ex}{$\gppr$}}

\usepackage{newtxtext,newtxmath}
\title[Dynamo activity in metal polluted white dwarfs]{Magnetic dynamos in white dwarfs -- II. Relating magnetism and pollution}

\author[M.R. Schreiber et al.]{
Matthias R. Schreiber$^{1,2}$\thanks{matthias.schreiber@usm.cl (MRS)},
Diogo Belloni$^{1,3}$,%\thanks{diogo.belloni@inpe.br (DB)},
Boris T. G\"ansicke$^{4}$, Steven G. Parsons$^5$
\\
$^1$Departamento de F\'isica, Universidad T\'ecnica Federico Santa Mar\'ia, Av. España 1680, Valpara\'iso, Chile \\
$^2$Millennium Nucleus for Planet Formation (NPF), Valpara\'iso, Chile \\
$^{3}$National Institute for Space Research, Av. dos Astronautas, 1758, 12227-010, S\~ao Jos\'e dos Campos, SP, Brazil\\
$^4$Department of Physics, University of Warwick, Coventry CV4 7AL, UK\\
$^5$Department of Physics and Astronomy, University of Sheffield, Sheffield S3 7RH, UK
}

\date{Accepted XXX. Received YYY; in original form ZZZ}

\pubyear{2015}

\begin{document}
\label{firstpage}
\pagerange{\pageref{firstpage}--\pageref{lastpage}}
\maketitle

\begin{abstract}
We investigate whether the recently suggested rotation and crystallization driven dynamo  can explain the apparent increase of magnetism in old metal polluted white dwarfs. We find that the effective temperature distribution of polluted magnetic white dwarfs is in agreement with most/all of them having a crystallizing core and increased rotational velocities are expected due to accretion of planetary material which is evidenced by the 
metal absorption lines. We conclude that a rotation and crystallization driven dynamo offers not only an explanation for the different occurrence rates of strongly magnetic white dwarfs in close binaries, but also for the high incidence of weaker magnetic fields in old metal polluted white dwarfs. 
\end{abstract}


%%%%%%%%%%%%%%%%%%%%%%%%%%%%%%%%%%%%%%%%%%%%%%%%%%

%%%%%%%%%%%%%%%%% BODY OF PAPER %%%%%%%%%%%%%%%%%%

\section{Introduction}

Metal absorption lines in the spectra of white dwarfs have been 
firmly established to result from the accretion of planetary material that survived the transformation of their 
host star into a white dwarf. This idea was first proposed by \citet{jura03-1} and later confirmed to be correct due to the detection of a transiting planetesimal in the process of tidal disintegration \citep{vanderburgetal15-1,gaensickeetal16-1}. 
This key evidence is complemented by the detection of dusty and gaseous debris disks \citep[e.g.][]{zuckerman+becklin87-1,gaensickeetal06-1}, a planetesimal that is possibly the core of a differentiated rocky planet 
\citep{manseretal19-1} and an evaporating planet in close orbit around a white dwarf
\citep{gaensickeetal19-1}. 

This growing observational evidence for the accretion of planetary material onto white dwarfs is in agreement with theoretical predictions. It is well established that planets can survive the transition of 
their host star into a white dwarf \citep[e.g.][]{villaver+livio09-1,roncoetal20-1}. Some of the surviving planetary systems are predicted 
to become unstable which can push especially lower mass planets or asteroids into highly eccentric 
orbits \citep{veras+gaensicke15-1,smallwoodetal18-1,maldonadoetal20-1}. 
Tidal forces may then destroy these objects which offers a consistent explanation for the observations of planetary material being accreted onto 25--50 per cent of all white dwarfs \citep{koesteretal14-1}. 

As noticed by \citet{hollandsetal15-1}, very old metal polluted white dwarfs are more frequently magnetic than younger systems. All magnetic metal polluted helium (DZ) white dwarfs 
and all but one known magnetic metal polluted hydrogen atmosphere white dwarfs (DAZ) 
have effective temperatures below 7500\,K \citep[][]{hollandsetal17-1,kawkaetal19-1,kawkaetal21-1}. 
In addition, 
three apparently isolated white dwarfs exhibiting Zeeman split {\em{emission}} lines which might be related to the existence of a conductive planet or planet core in a close orbit, cluster around temperatures of $\sim7500$K \citep[e.g.][]{lietal98-1,gaensickeetal20-1,waltersetal21-1}. 
This suggests that the accretion of planetary material  
and low temperatures may be linked to the generation of magnetic fields in white dwarfs.  

Inspired by the earlier work of \citet{isernetal17-1}, we recently proposed a dynamo similar to those operating
in planets and low-mass stars to explain the observed incidence of strongly magnetic white dwarfs in close binary stars \citep{schreiberetal21-1}. The main ingredients for this dynamo to work are that the white dwarf's core started to crystallize (which depending on the mass of the white dwarf generally occurs at white dwarf temperatures below $\simeq8000$\,K) and increased rotational velocities due to accretion.  
We here test the hypothesis that the increasing fraction of magnetic white dwarfs among cold 
metal polluted (and therefore accreting) white dwarfs could be produced by the same dynamo mechanism.  
We start with a brief review of the dynamo mechanism recently suggested to work in accreting white dwarfs in close binaries. 

\vspace{-0.25cm}

\section{The convective dynamo applied to white dwarfs}

The magnetic fields of planets 
and rapidly rotating low-mass stars, are generated by convection-driven dynamos \citep{christensenetal09-1}. 
The main ingredients for these dynamos to work are a strong density stratification, an extended convection zone, and rotation. A similar configuration can occur in cooling white dwarfs. 

As a carbon/oxygen white dwarf cools, the ions in the core begin to freeze in a lattice structure \citep{vanhorn68-1}, i.e. the white dwarf starts to crystallize.  
The phase diagram of the carbon–oxygen mixture is of the spindle form \citep[e.g.][]{horowitzetal10-1} and 
consequently, the solid phase becomes richer in oxygen and sinks while the carbon excess mixes with the outer liquid envelope 
which is redistributed by the Rayleigh–Taylor instability \citep[e.g.][]{isernetal97-1}. If the white dwarf is 
also rapidly rotating, the conditions are appropriate for
magnetic field generation through the convective dynamo. 

For planets and low-mass stars, the magnetic field strength can be derived 
from fundamental properties of a given object using
the convective energy flux scaling law \citep{christensenetal09-1}.   
Applying this scaling law to white dwarfs led to the prediction 
of field strengths below $\sim\,1$\,MG \citep{isernetal17-1}.
However, the field strength generated by the dynamo likely depends on the magnetic Prandtl number \citep{brandenburg14-1} which is not taken into account in the scaling law \citep{christensen+aubert06-1}. 
Given that the magnetic Prandtl number for crystallizing white dwarfs is orders of magnitude larger than that of planets and low-mass stars, 
the field strength generated by the dynamo is likely also much larger \citep{bovinoetal13-1}. For more details see \citet{schreiberetal21-1}.

According to measurements of activity and rotation in fully convective stars, the dynamo seems to saturate for 
Rossby numbers (rotation period divided by convective turnover time) below $0.1$
which might indicate that the generated field strength
becomes independent of the rotation rate for rapidly
spinning stars.  
For white dwarfs, the condition on the Rossby number translates to rotation periods of the order of seconds/minutes as the threshold for saturation. For slower rotation rates, magnetic fields might still be generated but should on average be weaker. 

Based on the reasonable assumption that strong magnetic fields can be generated if a crystallizing white dwarf is rotating 
in the saturated regime, \citet{schreiberetal21-1} explained the occurrence rates and characteristics of strongly magnetic white dwarfs in close binary stars. 
The very same mechanism may explain the large incidence of weaker magnetic fields in cool metal 
polluted white dwarfs if they have temperatures consistent with crystallizing cores and if the accretion of planetary material can significantly increase their rotation. 
In the next section we evaluate whether the distribution of effective temperatures of magnetic DAZ and DZ white dwarfs 
is consistent with them having crystallizing cores. 
\vspace{-0.25cm}
\section{Crystallizing cores in magnetic metal polluted white dwarfs}

\begin{figure}
  \begin{center}
     \includegraphics[width=0.99\linewidth]{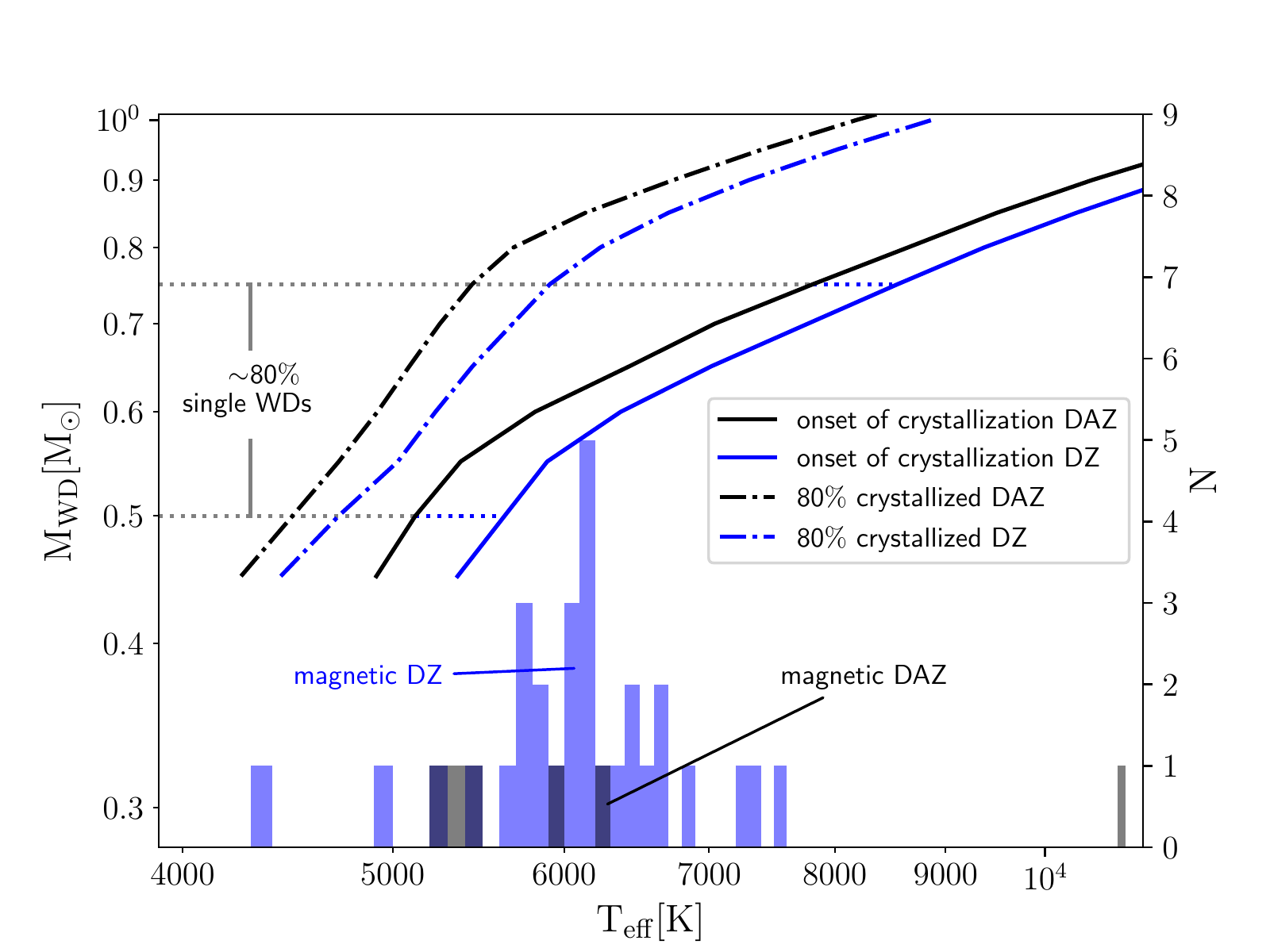}
  \end{center}
  \caption{The onset and 80 per cent completed crystallization temperatures for thin (blue) and thick (black) hydrogen atmospheres taken from \citet{bedardetal20-1}. For thin atmospheres (DZ) crystallization starts at slightly higher temperatures than for thick atmospheres (DAZ). Crystallization starts at effective 
  temperatures well above 10\,000\,K for 
  massive white dwarfs but for typical white dwarf masses in the range 5000--9000\,K. The currently known DZ (blue histogram) and DAZ (grey histogram) magnetic white dwarfs lie exactly in the range temperature range of crystallizing cores for typical white dwarf masses. The only exception is WD\,2105--820 which is hotter and (most likely and fittingly) more massive \citep{swanetal19-1}. }
  \label{FigCrys}
\end{figure}

The onset of crystallization in white dwarfs, a requirement for the dynamo mechanism proposed by \citet{schreiberetal21-1}, depends not only on the white dwarf effective temperature but also on the white dwarf mass. 
However, measuring the masses of magnetic white dwarfs is extremely challenging. The standard technique for non-magnetic DA white dwarfs, which make up the bulk of the white dwarf population \citep[e.g.][]{mccleeryetal20-1}, is to measure the effective temperature and surface gravity from fitting the Balmer lines and then to determine the white dwarf mass using a mass-radius relation. There is currently no theory for the simultaneous treatment of the Zeeman effect and Stark broadening, and therefore, this method cannot be applied to magnetic DA(Z) white dwarfs. 

An alternative method to derive white dwarf masses is to iteratively fit the photometric spectral energy distribution (SED, which, for a given temperature, is very sensitive to the radius) and the spectrum (where the relative strengths of absorption lines provide a handle on the temperature), and to subsequently derive the mass adopting a white dwarf mass-radius relation. However, this method requires the accurate knowledge of the distance to the white dwarf under analysis. Most cool and faint DZ white dwarfs in our sample have large parallax uncertainties in the available \textit{Gaia} data releases, and consequently, estimated masses are unreliable \citep{coutuetal19-1}. Additional complications arise from the line blanketing caused by the metals in the magnetic DZ white dwarfs considered here, introducing the detailed photospheric abundances as additional free fit parameters. Finally, magnetic fields may suppress or weaken convection, which will affect the temperature structure and emerging spectrum of magnetic white dwarfs \citep{gentile-fusilloetal18-1}, and it is currently unclear whether the SEDs of magnetic white dwarfs are better fitted with radiative or convective models.
Given the difficulties in determining the individual masses of the magnetic metal polluted white dwarfs, we here investigate whether the {\em distribution} of their effective temperatures is in agreement with their cores being in the process of crystallization. 

Largely based on earlier work \citep{hollandsetal17-1}, 
evidence is growing that the occurrence of magnetic fields in DZ white dwarfs increases with decreasing temperature \citep{kawkaetal21-1}. All magnetic DZ white dwarfs have effective temperatures below 7500\,K. At these low temperatures $\simeq40$ per cent of the known DZ white dwarfs are magnetic. In contrast, 
not a single magnetic DZ has been found in the temperature range $7500-10000$\,K.
The situation is very similar for DAZ white dwarfs among which  
the fraction of magnetic systems increases significantly with decreasing effective temperature and all but one magnetic DAZ white dwarfs are cooler than $\sim$7500\,K. 
A list of magnetic DAZ and DZ white dwarfs is provided in Table\,1 where we ignored some 
candidate magnetic systems suggested in \citet{kawka+vennes14-1} and \citet{coutuetal19-1} as well as the magnetic Balmer emission line white dwarfs discussed by \citet{gaensickeetal20-1}.  

Comparing the measured temperatures with those expected for crystallizing white dwarfs, we find good agreement 
with ongoing crystallization in their 
cores (see Fig.\,\ref{FigCrys}). 
Roughly 80 per cent of single white dwarfs can be found in the mass range of $0.5-0.75$\Msun\ \citep{tremblayetal16-1} and the distribution peaks at $\simeq0.6$\,\Msun.
All magnetic DZ and all but one magnetic DAZ white dwarfs are 
cooler than the temperature at which a 
0.75\Msun\ white dwarf starts to crystallize ($\sim8\,000$\,K) and roughly half of them are below the crystallization temperature 
of a 0.6\,\Msun\ white dwarf (see Fig.\,\ref{FigCrys}). 
Only the hottest ($\sim11\,000$\,K) metal polluted magnetic white dwarf, the DAZ white dwarf WD\,2105$-$820, seems to contradict this finding as its temperature is too high to be crystallizing if its mass was that of a typical 
0.5--0.75\Msun\ white dwarf. However, apart from being the hottest metal polluted magnetic white dwarf, WD\,2105$-$820 is also most likely the most massive one. \citet{swanetal19-1} estimated a white dwarf mass of $0.86$\,\Msun\ for which crystallization starts at much higher temperatures. The estimated mass is not high enough to reach the onset of crystallization but assuming an uncertainty of just $0.05$\Msun\ would fix this apparent disagreement (which would be completely consistent with the uncertainty provided for $\log\,g$). 
We conclude that indeed most/all metal polluted magnetic white dwarfs might have
passed the onset of crystallization in their cores. 

The remaining question is whether the accretion of planetary material could spin up the white dwarfs to reach significantly increased rotational velocities. 

\vspace{-0.25cm}

\begin{table}
%	\centering
	\caption{Currently known magnetic DZ and DAZ white dwarfs.}
	\label{tab:parameters}
	\begin{tabular}{lcccr} % four columns, alignment for each
		\hline
           Name &  Type & T$_{\mathrm{eff}}$ [K] & B$_S$ [MG] & reference\\
          		\hline
%WD\,1328+30 & DZ & 6000 & 0.65 & 0 \\
%SDSS\,J0902+3625 & DZ 6300 & 1.92$\pm$0.05 & 0 \\
%SDSS\,J1214
SDSS\,J0037--0525 & DZ & 5630$\pm$120 & 7.09$\pm$0.04 & 1 \\
SDSS\,J0107+2650 & DZ & 6190$\pm$140 & 3.37$\pm$0.07 & 1 \\
SDSS\,J0157+0033 & DZ & 6110$\pm$140 & 3.49 ± 0.05 & 1 \\
SDSS\,J0200+1646 & DZ & 5810$\pm$180 & 10.71 ± 0.07 & 1 \\
SDSS\,J0735+2057 & DZ & 6110$\pm$180 & 6.12 ± 0.06 & 1 \\
SDSS\,J0806+4058 & DZ & 6808$\pm$80 & 0.80 ± 0.03 & 1 \\
SDSS\,J0832+4109 & DZ & 6070$\pm$190 & 2.35 ± 0.11 & 1 \\
SDSS\,J0902+3625 & DZ & 6330$\pm$210 & 1.92 ± 0.05 & 1 \\
SDSS\,J0927+4931 & DZ & 6200$\pm$230 & 2.10 ± 0.09 & 1 \\
SDSS\,J1003-0031 & DZ & 5740$\pm$140 & 4.37 ± 0.05 & 1 \\
SDSS\,J1105+5006 & DZ & 7280$\pm$190 & 4.13 ± 0.11 & 1 \\
SDSS\,J1106+6737 & DZ & 6400$\pm$170 & 3.50 ± 0.09 & 1 \\
SDSS\,J1113+2751 & DZ & 6180$\pm$210 & 3.18 ± 0.09 & 1 \\
%WD\,1143+6615 & DZ & 
SDSS\,J1150+4533 & DZ & 5720$\pm$320 & 2.01$\pm$0.20 & 1 \\
%WD\,1150+4928 & DZ & 7210$\pm$160 & 1 \\
SDSS\,J1152+1605 & DZ & 6550$\pm$160 & 2.72$\pm$0.04  & 1 \\
SDSS\,J1214--0234 & DZ & 5210$\pm$100 & 2.12$\pm$0.03 & 1 \\
SDSS\,J1249+6514 & DZ & 7540$\pm$170 & 2.15$\pm$0.05 & 1 \\
SDSS\,J1330+3029 & DZ & 6100$\pm$60 & 0.57$\pm$0.04 & 1 \\
SDSS\,J1412+2836 & DZ & 4990$\pm$160 & 1.99$\pm$0.10 & 1 \\
SDSS\,J1536+4205 & DZ & 5800$\pm$140 & 9.59$\pm$0.04 & 1 \\
SDSS\,J1546+3009 & DZ & 6600$\pm$120 & 0.81$\pm$0.07 & 1 \\
SDSS\,J1651+4249 & DZ & 5710$\pm$200 & 3.12$\pm$0.28 & 1 \\
SDSS\,J2254+3031 & DZ & 5900$\pm$90 & 2.53$\pm$0.03 & 1 \\
SDSS\,J2325+0448 & DZ & 6020$\pm$100 & 6.56$\pm$0.09 & 1 \\
SDSS\,J2330+2805 & DZ & 6670$\pm$210 & 3.40$\pm$0.04 & 1 \\
%Tremblay et al. 2020 
WD\,1515+8230 & DZ & 4360$\pm$80 & 3.1$\pm$0.2 & 2 \\
%Bagnulo Landstreet 2019
WD\,0816--310 & DZ & 6436 & 0.092$\pm$0.001 & 3,4,5 \\%  (see also Kawka 2021) 
WD\,1009--184 & DZ & 6036 & $\gappr$0.3 & 3,6 \\
WD\,1532+129 & DZ & 5430 & $\gappr$0.3 & 3,4 \\
WD\,2138–-332 & DZ & 7399 & $\gappr$0.4 & 3,4 \\
%DAZ from Kawka et al 2019
WD\,0214--071 & DAZ & 5460$\pm$40 & 0.163$\pm$0.004 & 7 \\
WD\,0315--293 & DAZ & 5200$\pm$200 & 0.519$\pm$0.04 & 7,8,9 \\
WD\,0322--019 & DAZ & 5310$\pm$100 & 0.120 & 7,10,11 \\
WD\,1653+385 & DAZ & 5900 & 0.07 & 7,12 \\
WD\,2225+176 & DAZ & 6250$\pm$70 & 0.334$\pm$0.003 & 7,13 \\
WD\,2105$-$820 & DAZ & 10890$\pm$380 & $\simeq$0.043 & 7,14,15,16 \\
\hline
	\end{tabular}
%\begin{tablenotes}
\noindent
%\item 
References: 
(1) \citet{hollandsetal17-1}, 
(2) \citet{tremblayetal20-1}, 
(3) \citet{bagnulo+landstreet19-1}, 
(4) \citet{giammicheleetal12-1}
(5) \citet{kawkaetal21-1}
(6) \citet{subasavageetal17-1}
(7) \citet{kawkaetal19-1},
(8) \citet{kawka+vennes12-1},
(9) \citet{kawka+vennes11-1},
(10) \citet{farihietal11-1},
(11) \citet{farihietal18-1},
(12) \citet{zuckermanetal11-1}, 
(13) \citet{kawka+vennes14-1}
(14) \citet{koesteretal98-1}
(15) \citet{landstreetetal12-1}
(16) \citet{swanetal19-1}
%\end{tablenotes}
\end{table}

\section{Spin-up of the white dwarf}

The second ingredient for the proposed dynamo proposed 
to work is increased rotation of the white dwarf. 
It is well established that metal polluted white dwarfs accrete planetary debris that result from disintegrating planetesimals, asteroids, comets, or planets. 
During the accretion process, the white dwarf not only accretes mass, but also angular momentum \citep{alexanderetal20-1}. We here estimate whether the accreted angular momentum might sufficiently spin up the white dwarf to cause the generation of detectable magnetic fields.  

From a large sample of cold and old metal polluted white dwarfs \citet{hollandsetal17-1} 
derived a trend of decreasing accretion rates with an e-folding time of $\simeq1$\,Gyr. 
This observed long term trend, however, does not cover short episodes of much larger accretion rates that occur when rocky or even gas giant planets come too close to the white dwarfs. Observational evidence for such events has been provided recently \citep{manseretal19-1,gaensickeetal19-1}.
%who interpreted the accretion of volatile elements through a gaseous accretion disk as evidence for a close-in evaporating gas giant planet.  
In addition, the recently observed giant planet around WD\,1856+534 shows that even Jupiter mass planets may end up in close orbits around white dwarfs either due to common envelope evolution \citep{lagosetal21-1}, triple dynamics 
\citep{munoz+petrovich20-1}, or gravitational interactions 
\citep{maldonadoetal20-1}. 
Indeed, using configurations of planetary systems derived from observed systems, \citet{maldonadoetal20-1} showed that 
eccentricity pumping leading to the tidal disruption and/or evaporation of planets 
is by no means a rare event. For planetary systems consisting of five or six planets, star-planet collisions are expected in $\sim10$ per cent of the cases. 

The accretion of planetary material does not only lead to metal absorption lines in the spectra of white dwarfs, accretion of mass is accompanied by the accretion of angular momentum. The material from destroyed planetesimals, asteroids, comets, and planets accumulates in a Keplerian disk around the white dwarf and is then slowly accreted onto the white dwarf.  
\citet{kingetal91-1} derived the angular momentum balance equation for accreting white dwarfs in cataclysmic variables
and found
\begin{equation}
I\,\frac{{\rm d}\omega}{{\rm d}t} \ = \ 
\alpha \, ( \, -\dot{M}_{2} \, ) \, \left( G \, M_{\rm WD} \, R_{\rm WD} \, \right)^{1/2} \ + \ \left( 1 + \epsilon \right) \left( \dot{M}_{2} \, \eta \, R_{\rm WD}^2 \, \omega \right),
\label{Eq:spinup}
\end{equation}
%\
%
\noindent
where $\omega$ is the white dwarf spin, $I$ its moment of inertia, $G$ the gravitational constant, $\dot{M}_2$ the mass transfer rate averaged over nova cycles, and $M_{\rm WD}$ and $R_{\rm WD}$ are the white dwarf mass and radius.
The first term on the right-hand side of the equation corresponds to the spin-up due to accretion and the parameter $0\leq\alpha\leq1$ \citep[added by][]{schreiberetal21-1} represents the spin-up efficiency. 
The second term represents the spin-down due to material leaving the white dwarf which can be ignored in this work 
as nova eruptions are generally 
not expected ($\epsilon=-1.0$) given the small amount of accreted hydrogen (the accretion of an entire Jupiter mass planet could be a rare exception).  
We solved the non-homogeneous differential equation~\ref{Eq:spinup}
as in \citet{schreiberetal21-1} for total accreted mass (over a time span of several Gyr) of $10^{-6}-10^{-3}$\,\Msun\ which roughly covers the mass range from Earth to Jupiter. 

\begin{figure}
  \begin{center}
     \includegraphics[width=0.99\linewidth]{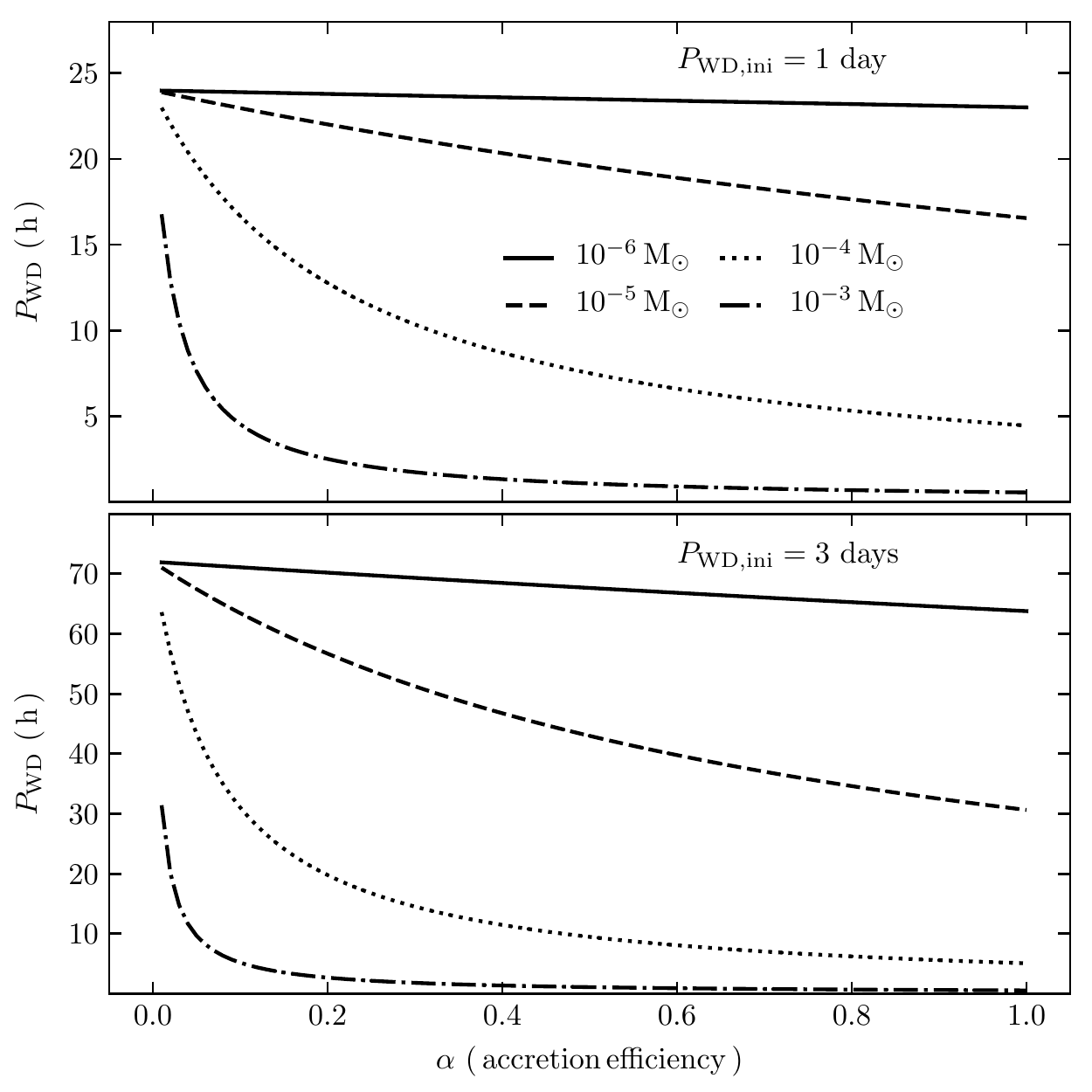}
  \end{center}
  \caption{Final white dwarf spin period ($P_{\rm WD}$) against the spin-up efficiency parameter ($\alpha$), for four different accreted masses, in \Msun, namely $10^{-6}$, $10^{-5}$, $10^{-4}$ and $10^{-3}$ and two initial spin periods. With the exception of accreted masses $\lesssim10^{-5}$, the white dwarf spin periods become significantly shorter (minutes to a few hours) for a large range of efficiencies.}
  \label{FigSpinUp}
\end{figure}

We find that the accretion of planetary material in polluted white dwarfs can easily 
lead to significantly shorter spin periods. 
Figure\,\ref{FigSpinUp} shows the spin-up of a $0.6$\,\Msun\ white dwarf for different total accretion rates 
as a function of the spin-up efficiency. For accreted masses exceeding $10^{-5}$\Msun~we find rotation periods ranging from several minutes, as recently observed \citep{redingetal20-1}, to a few hours, much shorter than the assumed initial rotation rate of 1--3\,days \citep{hermesetal17-1}. 
At the same time, the spin periods reached remain longer than the spin period estimated by \citet{isernetal17-1} 
for saturation of the dynamo which could be reached
due to the accretion of significantly more mass in Cataclysmic Variables (CVs). 
One would therefore expect the generated magnetic fields 
in DZ and DAZ white dwarfs to be weaker than the up to several 100\,MG 
fields of strongly magnetic white dwarfs in CVs which is clearly the case
(see Table\,1). 

We conclude that both conditions for the generation of magnetic fields due to a crystallization and rotation driven dynamo are most likely fulfilled in DZ and DAZ white dwarfs. 
To the best of our knowledge, no other scenario suggested for the generation of magnetic fields in white dwarfs offers an explanation for the increased incidence of magnetism in old metal polluted white dwarfs. 

However, in the absence of a scaling law that takes into account the dependence on the magnetic Prandtl number, let alone detailed simulations of the dynamo in white dwarfs, we admit that the presented arguments are reasonable but phenomenological. 
More detailed theoretical investigations as well as a representative sample of observed magnetic DZ and DAZ white dwarfs are clearly required to further test the outlined scenario. 

\vspace{-0.25cm}

\section{Predictions to be tested} 

At first glance, one might think that calculating the 
temperature distribution of crystallizing white dwarfs
by combining stellar evolution codes and white dwarf cooling tracks could be an easy way to confront the observed temperature distribution with model predictions. 
However, such a comparison would currently represent a rather futile exercise as the distributions of the effective temperatures of the currently available 
sample of DAZ and DZ white dwarfs is not only very likely biased towards hot white dwarfs but also incomplete. 
In addition, making reliable model predictions for the temperature distribution of crystallizing metal polluted white dwarfs is currently hardly possible as the occurrence rate of planetary systems around the progenitor stars and its exact dependence on the stellar mass, planet separation, planet mass, and stellar metallicity are, despite recent progress \citep[e.g.][]{fischer+valenti05-1,muldersetal15-1}, not well known. 
Therefore, instead of performing such a comparison, we here investigate whether the suggested dynamo scenario leads to predictions that can be observationally tested in the near future. 

To that end we calculated the temperature distribution of crystallizing DZ and DAZ white dwarfs for solar metallicity and assuming a constant probability for the existence of planetary systems up to masses of the white dwarf progenitor of $3.0$\Msun. As the planet occurrence rate seems to drop for larger stellar masses \citep[e.g.][]{muldersetal15-1}, we assumed that planets do not form around more massive stars. 
We furthermore assumed 
a constant star formation rate, an age of the Galactic disk of 10 Gyr, and an initial mass function $\propto\,M^{-2.3}$. 
We used the the single star evolution code written by \citet{hurleyetal00-1} and the white dwarf cooling models of \citet{bedardetal20-1}.

\begin{figure}
  \begin{center}
     \includegraphics[width=0.99\linewidth]{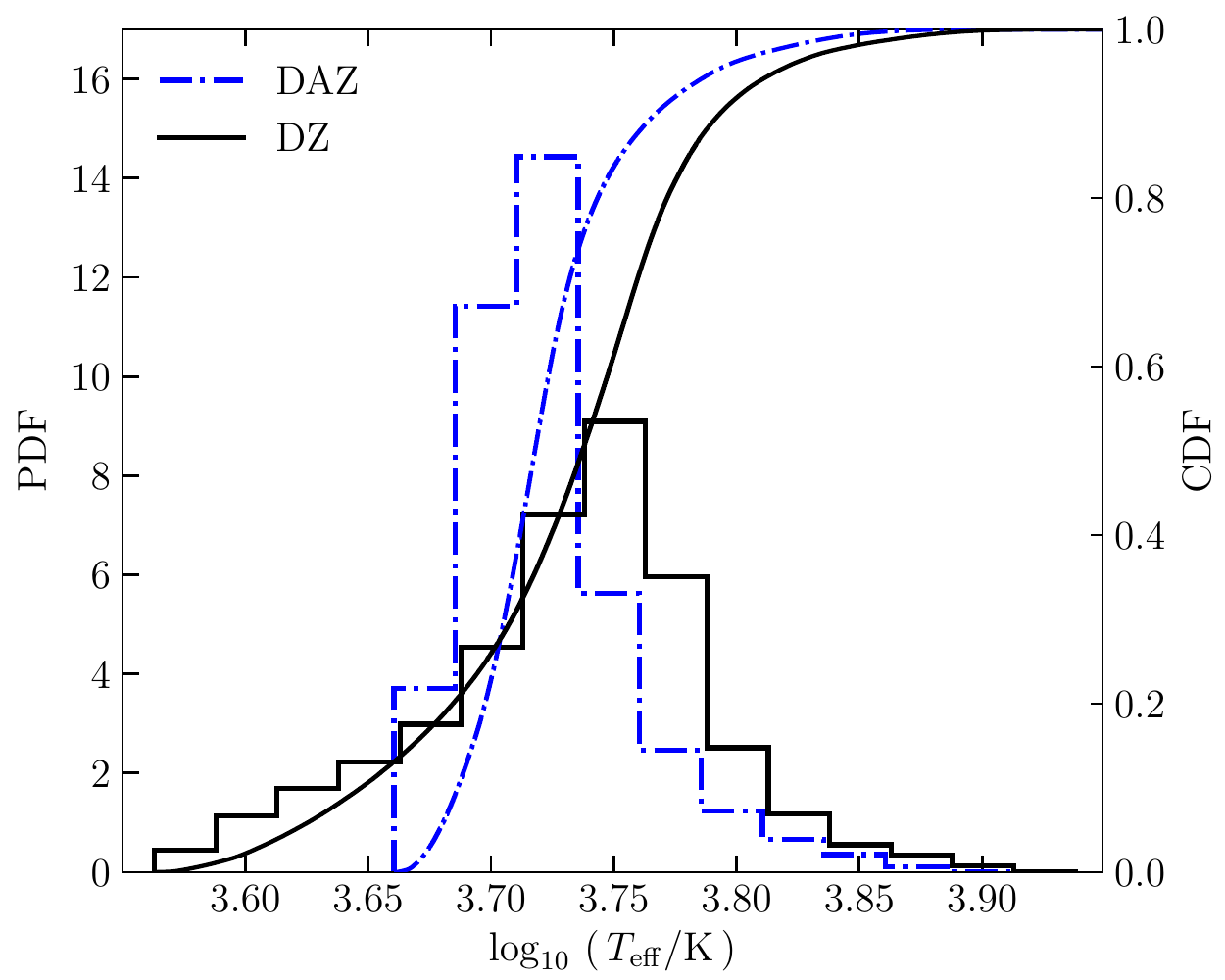}
  \end{center}
  \caption{Predicted temperature distributions of crystallizing DAZ and DZ white dwarfs assuming solar metallicity and a constant occurrence rate of planetary systems around progenitor stars less massive then 3.0\Msun. 
  The predicted distribution of temperatures of DZ white dwarfs is much broader and peaks at higher temperatures than the observed DZ sample. }
  \label{FigTeff}
\end{figure}

The resulting temperature distributions are shown in Fig.\,\ref{FigTeff}. 
As the onset of crystallization for DZ white dwarfs occurs at higher temperatures, their distribution peaks at a slightly higher temperature. Given that in addition the late cooling of DZ white dwarfs is faster, their predicted distribution also extends more towards cooler white dwarfs. 
While the detailed shape of both distributions depends on the assumed occurrence rate of planets as a function of stellar mass and metallicity, the predicted differences 
should be present as long as both samples suffer from the same observational biases and their progenitor stars have the same planet occurrence rates. 

The small number of magnetic DAZ and DZ white dwarfs currently known does not allow us to asses whether the underlying temperature distributions are different. However, the ongoing SDSS\,-V survey will provide spectra of $\sim200\,000$ white dwarfs, many of which will be metal polluted and a significant fraction of the latter will be magnetic. 
Thus, the required large sample of magnetic DZ and DAZ white dwarfs is likely to be established within the next few years. 

A second and obvious prediction of our model is that the rotation periods of magnetic DZ and DAZ white dwarfs should be on average shorter than those of non-magnetic white dwarfs with crystallizing cores. 
Fortunately, the rotation periods of magnetic white dwarfs can be measured 
through either photometric variability \citep[][]{brinkworthetal13-1} 
or circular spectropolarimetry of spectral lines \citep{bagnulo+landstreet19-1}
and thus the prediction can be tested. 
Interestingly, the three cold magnetic white dwarfs showing Balmer emission and potentially hosting a conductive planet or planetary core \citep{gaensickeetal20-1}, clearly show reduced spin periods which fits with the scenario outlined here.

Our scenario also predicts that a certain fraction of crystallizing white dwarfs currently not showing metal absorption lines should have increased rotation rates and be magnetic because accretion that spun up the white dwarf has occurred in the past. The cool magnetic white dwarfs found in the 20\,pc sample  \citep{bagnulo+landstreet20-1} 
could be such systems. 

Finally, another mechanism must be responsible for the magnetic fields observed in some hot white dwarfs and this alternative process must prevent the accretion of planetary debris (given the absence of hot magnetic metal polluted white dwarfs). 
Double white dwarf mergers 
could be such a mechanism, as planetary material is certainly unlikely to survive two phases of mass transfer.

\section{Conclusion}
\citet{schreiberetal21-1}
recently suggested a crystallisation and rotation driven dynamo for the origin of strong magnetic 
fields of white dwarfs in close binary stars. We here investigated whether the same mechanism might be responsible for the surprising increase of magnetism in metal polluted white dwarfs when they have cooled to temperatures $\lappr7500$\,K. We found that the temperature distribution of metal polluted magnetic DAZ and DZ white dwarfs is consistent with them having crystallizing cores and that the accretion of planetary material can spin up the white dwarf's rotation to periods ranging from minutes to a hours depending on the total amount of accreted material. 
Thus, the suggested dynamo represents a promising model for explaining the magnetic fields of old DA/DAZ white dwarfs. 

\vspace{-0.25cm}

\section*{Acknowledgements}
MRS acknowledges support from Fondecyt (grant 1181404) 
and ANID, -- Millennium Science Initiative Program -- NCN19\_171.
DB was supported by FAPESP, {grant \#2017/14289-3}
%S\~ao Paulo Research Foundation (FAPESP) 
and ESO/Gobierno de Chile. BTG was
supported by a Leverhulme Research Fellowship and the UK STFC
grant ST/T000406/1. SGP acknowledges the support of an STFC Ernest Rutherford Fellowship. 
%%%%%%%%%%%%%%%%%%%%%%%%%%%%%%%%%%%%%%%%%%%%%%%%%%

\vspace{-0.25cm}

\section*{Data Availability}
The simulated data will be provided upon request.

\vspace{-0.25cm}


%\bibliographystyle{mnras}
%\bibliography{mag_planet} % if your bibtex file is called example.bib

\begin{thebibliography}{}
\makeatletter
\relax
\def\mn@urlcharsother{\let\do\@makeother \do\$\do\&\do\#\do\^\do\_\do\%\do\~}
\def\mn@doi{\begingroup\mn@urlcharsother \@ifnextchar [ {\mn@doi@}
  {\mn@doi@[]}}
\def\mn@doi@[#1]#2{\def\@tempa{#1}\ifx\@tempa\@empty \href
  {http://dx.doi.org/#2} {doi:#2}\else \href {http://dx.doi.org/#2} {#1}\fi
  \endgroup}
\def\mn@eprint#1#2{\mn@eprint@#1:#2::\@nil}
\def\mn@eprint@arXiv#1{\href {http://arxiv.org/abs/#1} {{\tt arXiv:#1}}}
\def\mn@eprint@dblp#1{\href {http://dblp.uni-trier.de/rec/bibtex/#1.xml}
  {dblp:#1}}
\def\mn@eprint@#1:#2:#3:#4\@nil{\def\@tempa {#1}\def\@tempb {#2}\def\@tempc
  {#3}\ifx \@tempc \@empty \let \@tempc \@tempb \let \@tempb \@tempa \fi \ifx
  \@tempb \@empty \def\@tempb {arXiv}\fi \@ifundefined
  {mn@eprint@\@tempb}{\@tempb:\@tempc}{\expandafter \expandafter \csname
  mn@eprint@\@tempb\endcsname \expandafter{\@tempc}}}

\bibitem[\protect\citeauthoryear{{Bagnulo} \& {Landstreet}}{{Bagnulo} \&
  {Landstreet}}{2019}]{bagnulo+landstreet19-1}
{Bagnulo} S.,  {Landstreet} J.~D.,  2019, \mn@doi [\aap]
  {10.1051/0004-6361/201936068}, \href
  {https://ui.adsabs.harvard.edu/abs/2019A&A...630A..65B} {630, A65}

\bibitem[\protect\citeauthoryear{{Bagnulo} \& {Landstreet}}{{Bagnulo} \&
  {Landstreet}}{2020}]{bagnulo+landstreet20-1}
{Bagnulo} S.,  {Landstreet} J.~D.,  2020, \mn@doi [\aap]
  {10.1051/0004-6361/202038565}, \href
  {https://ui.adsabs.harvard.edu/abs/2020A&A...643A.134B} {643, A134}

\bibitem[\protect\citeauthoryear{{B{\'e}dard}, {Bergeron}, {Brassard}  \&
  {Fontaine}}{{B{\'e}dard} et~al.}{2020}]{bedardetal20-1}
{B{\'e}dard} A.,  {Bergeron} P.,  {Brassard} P.,   {Fontaine} G.,  2020,
  \mn@doi [\apj] {10.3847/1538-4357/abafbe}, \href
  {https://ui.adsabs.harvard.edu/abs/2020ApJ...901...93B} {901, 93}

\bibitem[\protect\citeauthoryear{{Bovino}, {Schleicher}  \& {Schober}}{{Bovino}
  et~al.}{2013}]{bovinoetal13-1}
{Bovino} S.,  {Schleicher} D.~R.~G.,   {Schober} J.,  2013, \mn@doi [New
  Journal of Phys.] {10.1088/1367-2630/15/1/013055}, \href
  {https://ui.adsabs.harvard.edu/abs/2013NJPh...15a3055B} {15, 013055}

\bibitem[\protect\citeauthoryear{{Brandenburg}}{{Brandenburg}}{2014}]{brandenburg14-1}
{Brandenburg} A.,  2014, \mn@doi [ApJ] {10.1088/0004-637X/791/1/12}, \href
  {https://ui.adsabs.harvard.edu/abs/2014ApJ...791...12B} {791, 12}

\bibitem[\protect\citeauthoryear{{Brinkworth}, {Burleigh}, {Lawrie}, {Marsh}
  \& {Knigge}}{{Brinkworth} et~al.}{2013}]{brinkworthetal13-1}
{Brinkworth} C.~S.,  {Burleigh} M.~R.,  {Lawrie} K.,  {Marsh} T.~R.,   {Knigge}
  C.,  2013, \mn@doi [\apj] {10.1088/0004-637X/773/1/47}, \href
  {https://ui.adsabs.harvard.edu/abs/2013ApJ...773...47B} {773, 47}

\bibitem[\protect\citeauthoryear{{Christensen} \& {Aubert}}{{Christensen} \&
  {Aubert}}{2006}]{christensen+aubert06-1}
{Christensen} U.~R.,  {Aubert} J.,  2006, \mn@doi [Geophys. Journal
  International] {10.1111/j.1365-246X.2006.03009.x}, \href
  {https://ui.adsabs.harvard.edu/abs/2006GeoJI.166...97C} {166, 97}

\bibitem[\protect\citeauthoryear{{Christensen}, {Holzwarth}  \&
  {Reiners}}{{Christensen} et~al.}{2009}]{christensenetal09-1}
{Christensen} U.~R.,  {Holzwarth} V.,   {Reiners} A.,  2009, \mn@doi [\nat]
  {10.1038/nature07626}, \href
  {https://ui.adsabs.harvard.edu/abs/2009Natur.457..167C} {457, 167}

\bibitem[\protect\citeauthoryear{{Coutu}, {Dufour}, {Bergeron}, {Blouin},
  {Loranger}, {Allard}  \& {Dunlap}}{{Coutu} et~al.}{2019}]{coutuetal19-1}
{Coutu} S.,  {Dufour} P.,  {Bergeron} P.,  {Blouin} S.,  {Loranger} E.,
  {Allard} N.~F.,   {Dunlap} B.~H.,  2019, \mn@doi [\apj]
  {10.3847/1538-4357/ab46b9}, \href
  {https://ui.adsabs.harvard.edu/abs/2019ApJ...885...74C} {885, 74}

\bibitem[\protect\citeauthoryear{{Farihi}, {Dufour}, {Napiwotzki}  \&
  {Koester}}{{Farihi} et~al.}{2011}]{farihietal11-1}
{Farihi} J.,  {Dufour} P.,  {Napiwotzki} R.,   {Koester} D.,  2011, \mn@doi
  [\mnras] {10.1111/j.1365-2966.2011.18325.x}, \href
  {https://ui.adsabs.harvard.edu/abs/2011MNRAS.413.2559F} {413, 2559}

\bibitem[\protect\citeauthoryear{{Farihi} et~al.,}{{Farihi}
  et~al.}{2018}]{farihietal18-1}
{Farihi} J.,  et~al., 2018, \mn@doi [\mnras] {10.1093/mnras/stx2664}, \href
  {https://ui.adsabs.harvard.edu/abs/2018MNRAS.474..947F} {474, 947}

\bibitem[\protect\citeauthoryear{{Fischer} \& {Valenti}}{{Fischer} \&
  {Valenti}}{2005}]{fischer+valenti05-1}
{Fischer} D.~A.,  {Valenti} J.,  2005, \mn@doi [\apj] {10.1086/428383}, \href
  {https://ui.adsabs.harvard.edu/abs/2005ApJ...622.1102F} {622, 1102}

\bibitem[\protect\citeauthoryear{{G{\"a}nsicke}, {Marsh}, {Southworth}  \&
  {Rebassa-Mansergas}}{{G{\"a}nsicke} et~al.}{2006}]{gaensickeetal06-1}
{G{\"a}nsicke} B.~T.,  {Marsh} T.~R.,  {Southworth} J.,   {Rebassa-Mansergas}
  A.,  2006, \mn@doi [Science] {10.1126/science.1135033}, \href
  {https://ui.adsabs.harvard.edu/abs/2006Sci...314.1908G} {314, 1908}

\bibitem[\protect\citeauthoryear{{G{\"a}nsicke} et~al.,}{{G{\"a}nsicke}
  et~al.}{2016}]{gaensickeetal16-1}
{G{\"a}nsicke} B.~T.,  et~al., 2016, \mn@doi [\apjl]
  {10.3847/2041-8205/818/1/L7}, \href
  {https://ui.adsabs.harvard.edu/abs/2016ApJ...818L...7G} {818, L7}

\bibitem[\protect\citeauthoryear{{G{\"a}nsicke}, {Schreiber}, {Toloza},
  {Fusillo}, {Koester}  \& {Manser}}{{G{\"a}nsicke}
  et~al.}{2019}]{gaensickeetal19-1}
{G{\"a}nsicke} B.~T.,  {Schreiber} M.~R.,  {Toloza} O.,  {Fusillo} N. P.~G.,
  {Koester} D.,   {Manser} C.~J.,  2019, \mn@doi [\nat]
  {10.1038/s41586-019-1789-8}, \href
  {https://ui.adsabs.harvard.edu/abs/2019Natur.576...61G} {576, 61}

\bibitem[\protect\citeauthoryear{{G{\"a}nsicke}, {Rodr{\'\i}guez-Gil}, {Gentile
  Fusillo}, {Inight}, {Schreiber}, {Pala}  \& {Tremblay}}{{G{\"a}nsicke}
  et~al.}{2020}]{gaensickeetal20-1}
{G{\"a}nsicke} B.~T.,  {Rodr{\'\i}guez-Gil} P.,  {Gentile Fusillo} N.~P.,
  {Inight} K.,  {Schreiber} M.~R.,  {Pala} A.~F.,   {Tremblay} P.-E.,  2020,
  \mn@doi [\mnras] {10.1093/mnras/staa2969}, \href
  {https://ui.adsabs.harvard.edu/abs/2020MNRAS.499.2564G} {499, 2564}

\bibitem[\protect\citeauthoryear{{Gentile Fusillo}, {Tremblay}, {Jordan},
  {G{\"a}nsicke}, {Kalirai}  \& {Cummings}}{{Gentile Fusillo}
  et~al.}{2018}]{gentile-fusilloetal18-1}
{Gentile Fusillo} N.~P.,  {Tremblay} P.~E.,  {Jordan} S.,  {G{\"a}nsicke}
  B.~T.,  {Kalirai} J.~S.,   {Cummings} J.,  2018, \mn@doi [\mnras]
  {10.1093/mnras/stx2584}, \href
  {https://ui.adsabs.harvard.edu/abs/2018MNRAS.473.3693G} {473, 3693}

\bibitem[\protect\citeauthoryear{{Giammichele}, {Bergeron}  \&
  {Dufour}}{{Giammichele} et~al.}{2012}]{giammicheleetal12-1}
{Giammichele} N.,  {Bergeron} P.,   {Dufour} P.,  2012, \mn@doi [\apjs]
  {10.1088/0067-0049/199/2/29}, \href
  {https://ui.adsabs.harvard.edu/abs/2012ApJS..199...29G} {199, 29}

\bibitem[\protect\citeauthoryear{{Hermes} et~al.,}{{Hermes}
  et~al.}{2017}]{hermesetal17-1}
{Hermes} J.~J.,  et~al., 2017, \mn@doi [\apjs] {10.3847/1538-4365/aa8bb5},
  \href {https://ui.adsabs.harvard.edu/abs/2017ApJS..232...23H} {232, 23}

\bibitem[\protect\citeauthoryear{{Hollands}, {G{\"a}nsicke}  \&
  {Koester}}{{Hollands} et~al.}{2015}]{hollandsetal15-1}
{Hollands} M.~A.,  {G{\"a}nsicke} B.~T.,   {Koester} D.,  2015, \mn@doi
  [\mnras] {10.1093/mnras/stv570}, \href
  {https://ui.adsabs.harvard.edu/abs/2015MNRAS.450..681H} {450, 681}

\bibitem[\protect\citeauthoryear{{Hollands}, {Koester}, {Alekseev}, {Herbert}
  \& {G{\"a}nsicke}}{{Hollands} et~al.}{2017}]{hollandsetal17-1}
{Hollands} M.~A.,  {Koester} D.,  {Alekseev} V.,  {Herbert} E.~L.,
  {G{\"a}nsicke} B.~T.,  2017, \mn@doi [\mnras] {10.1093/mnras/stx250}, \href
  {https://ui.adsabs.harvard.edu/abs/2017MNRAS.467.4970H} {467, 4970}

\bibitem[\protect\citeauthoryear{{Horowitz}, {Schneider}  \&
  {Berry}}{{Horowitz} et~al.}{2010}]{horowitzetal10-1}
{Horowitz} C.~J.,  {Schneider} A.~S.,   {Berry} D.~K.,  2010, \mn@doi [Phys.
  Rev. Lett.] {10.1103/PhysRevLett.104.231101}, \href
  {https://ui.adsabs.harvard.edu/abs/2010PhRvL.104w1101H} {104, 231101}

\bibitem[\protect\citeauthoryear{{Hurley}, {Pols}  \& {Tout}}{{Hurley}
  et~al.}{2000}]{hurleyetal00-1}
{Hurley} J.~R.,  {Pols} O.~R.,   {Tout} C.~A.,  2000, \mn@doi [\mnras]
  {10.1046/j.1365-8711.2000.03426.x}, \href
  {https://ui.adsabs.harvard.edu/abs/2000MNRAS.315..543H} {315, 543}

\bibitem[\protect\citeauthoryear{{Isern}, {Mochkovitch}, {Garc{\'\i}a-Berro}
  \& {Hernanz}}{{Isern} et~al.}{1997}]{isernetal97-1}
{Isern} J.,  {Mochkovitch} R.,  {Garc{\'\i}a-Berro} E.,   {Hernanz} M.,  1997,
  \mn@doi [ApJ] {10.1086/304425}, \href
  {https://ui.adsabs.harvard.edu/abs/1997ApJ...485..308I} {485, 308}

\bibitem[\protect\citeauthoryear{{Isern}, {Garc{\'\i}a-Berro}, {K{\"u}lebi}  \&
  {Lor{\'e}n-Aguilar}}{{Isern} et~al.}{2017}]{isernetal17-1}
{Isern} J.,  {Garc{\'\i}a-Berro} E.,  {K{\"u}lebi} B.,   {Lor{\'e}n-Aguilar}
  P.,  2017, \mn@doi [ApJ] {10.3847/2041-8213/aa5eae}, \href
  {https://ui.adsabs.harvard.edu/abs/2017ApJ...836L..28I} {836, L28}

\bibitem[\protect\citeauthoryear{{Jura}}{{Jura}}{2003}]{jura03-1}
{Jura} M.,  2003, \mn@doi [\apjl] {10.1086/374036}, \href
  {https://ui.adsabs.harvard.edu/abs/2003ApJ...584L..91J} {584, L91}

\bibitem[\protect\citeauthoryear{{Kawka} \& {Vennes}}{{Kawka} \&
  {Vennes}}{2011}]{kawka+vennes11-1}
{Kawka} A.,  {Vennes} S.,  2011, \mn@doi [\aap] {10.1051/0004-6361/201117078},
  \href {https://ui.adsabs.harvard.edu/abs/2011A&A...532A...7K} {532, A7}

\bibitem[\protect\citeauthoryear{{Kawka} \& {Vennes}}{{Kawka} \&
  {Vennes}}{2012}]{kawka+vennes12-1}
{Kawka} A.,  {Vennes} S.,  2012, \mn@doi [\aap] {10.1051/0004-6361/201118210},
  \href {https://ui.adsabs.harvard.edu/abs/2012A&A...538A..13K} {538, A13}

\bibitem[\protect\citeauthoryear{{Kawka} \& {Vennes}}{{Kawka} \&
  {Vennes}}{2014}]{kawka+vennes14-1}
{Kawka} A.,  {Vennes} S.,  2014, \mn@doi [\mnras] {10.1093/mnrasl/slu004},
  \href {https://ui.adsabs.harvard.edu/abs/2014MNRAS.439L..90K} {439, L90}

\bibitem[\protect\citeauthoryear{{Kawka}, {Vennes}, {Ferrario}  \&
  {Paunzen}}{{Kawka} et~al.}{2019}]{kawkaetal19-1}
{Kawka} A.,  {Vennes} S.,  {Ferrario} L.,   {Paunzen} E.,  2019, \mn@doi
  [\mnras] {10.1093/mnras/sty3048}, \href
  {https://ui.adsabs.harvard.edu/abs/2019MNRAS.482.5201K} {482, 5201}

\bibitem[\protect\citeauthoryear{{Kawka}, {Vennes}, {Allard}, {Leininger}  \&
  {Gad{\'e}a}}{{Kawka} et~al.}{2021}]{kawkaetal21-1}
{Kawka} A.,  {Vennes} S.,  {Allard} N.~F.,  {Leininger} T.,   {Gad{\'e}a}
  F.~X.,  2021, \mn@doi [\mnras] {10.1093/mnras/staa3421}, \href
  {https://ui.adsabs.harvard.edu/abs/2021MNRAS.500.2732K} {500, 2732}

\bibitem[\protect\citeauthoryear{{King}, {Regev}  \& {Wynn}}{{King}
  et~al.}{1991}]{kingetal91-1}
{King} A.~R.,  {Regev} O.,   {Wynn} G.~A.,  1991, \mn@doi [\mnras]
  {10.1093/mnras/251.1.30P}, \href
  {https://ui.adsabs.harvard.edu/abs/1991MNRAS.251P..30K} {251, 30P}

\bibitem[\protect\citeauthoryear{{Koester}, {Dreizler}, {Weidemann}  \&
  {Allard}}{{Koester} et~al.}{1998}]{koesteretal98-1}
{Koester} D.,  {Dreizler} S.,  {Weidemann} V.,   {Allard} N.~F.,  1998, \aap,
  \href {https://ui.adsabs.harvard.edu/abs/1998A&A...338..612K} {338, 612}

\bibitem[\protect\citeauthoryear{{Koester}, {G{\"a}nsicke}  \&
  {Farihi}}{{Koester} et~al.}{2014}]{koesteretal14-1}
{Koester} D.,  {G{\"a}nsicke} B.~T.,   {Farihi} J.,  2014, \mn@doi [\aap]
  {10.1051/0004-6361/201423691}, \href
  {https://ui.adsabs.harvard.edu/abs/2014A&A...566A..34K} {566, A34}

\bibitem[\protect\citeauthoryear{{Lagos}, {Schreiber}, {Zorotovic},
  {G{\"a}nsicke}, {Ronco}  \& {Hamers}}{{Lagos} et~al.}{2021}]{lagosetal21-1}
{Lagos} F.,  {Schreiber} M.~R.,  {Zorotovic} M.,  {G{\"a}nsicke} B.~T.,
  {Ronco} M.~P.,   {Hamers} A.~S.,  2021, \mn@doi [\mnras]
  {10.1093/mnras/staa3703}, \href
  {https://ui.adsabs.harvard.edu/abs/2021MNRAS.501..676L} {501, 676}

\bibitem[\protect\citeauthoryear{{Landstreet}, {Bagnulo}, {Valyavin},
  {Fossati}, {Jordan}, {Monin}  \& {Wade}}{{Landstreet}
  et~al.}{2012}]{landstreetetal12-1}
{Landstreet} J.~D.,  {Bagnulo} S.,  {Valyavin} G.~G.,  {Fossati} L.,  {Jordan}
  S.,  {Monin} D.,   {Wade} G.~A.,  2012, \mn@doi [\aap]
  {10.1051/0004-6361/201219829}, \href
  {https://ui.adsabs.harvard.edu/abs/2012A&A...545A..30L} {545, A30}

\bibitem[\protect\citeauthoryear{{Li}, {Ferrario}  \& {Wickramasinghe}}{{Li}
  et~al.}{1998}]{lietal98-1}
{Li} J.,  {Ferrario} L.,   {Wickramasinghe} D.,  1998, \mn@doi [\apjl]
  {10.1086/311546}, \href
  {https://ui.adsabs.harvard.edu/abs/1998ApJ...503L.151L} {503, L151}

\bibitem[\protect\citeauthoryear{{Maldonado}, {Villaver}, {Mustill},
  {Ch{\'a}vez}  \& {Bertone}}{{Maldonado} et~al.}{2020}]{maldonadoetal20-1}
{Maldonado} R.~F.,  {Villaver} E.,  {Mustill} A.~J.,  {Ch{\'a}vez} M.,
  {Bertone} E.,  2020, arXiv e-prints, \href
  {https://ui.adsabs.harvard.edu/abs/2020arXiv201011403M} {p. arXiv:2010.11403}

\bibitem[\protect\citeauthoryear{Manser et~al.,}{Manser
  et~al.}{2019}]{manseretal19-1}
Manser C.~J.,  et~al., 2019, \mn@doi [Science] {10.1126/science.aat5330}, 364,
  66

\bibitem[\protect\citeauthoryear{{McCleery} et~al.,}{{McCleery}
  et~al.}{2020}]{mccleeryetal20-1}
{McCleery} J.,  et~al., 2020, \mn@doi [\mnras] {10.1093/mnras/staa2030}, \href
  {https://ui.adsabs.harvard.edu/abs/2020MNRAS.499.1890M} {499, 1890}

\bibitem[\protect\citeauthoryear{{Mu{\~n}oz} \& {Petrovich}}{{Mu{\~n}oz} \&
  {Petrovich}}{2020}]{munoz+petrovich20-1}
{Mu{\~n}oz} D.~J.,  {Petrovich} C.,  2020, \mn@doi [\apjl]
  {10.3847/2041-8213/abc564}, \href
  {https://ui.adsabs.harvard.edu/abs/2020ApJ...904L...3M} {904, L3}

\bibitem[\protect\citeauthoryear{{Mulders}, {Pascucci}  \& {Apai}}{{Mulders}
  et~al.}{2015}]{muldersetal15-1}
{Mulders} G.~D.,  {Pascucci} I.,   {Apai} D.,  2015, \mn@doi [\apj]
  {10.1088/0004-637X/814/2/130}, \href
  {https://ui.adsabs.harvard.edu/abs/2015ApJ...814..130M} {814, 130}

\bibitem[\protect\citeauthoryear{{Reding}, {Hermes}, {Vanderbosch}, {Dennihy},
  {Kaiser}, {Mace}, {Dunlap}  \& {Clemens}}{{Reding}
  et~al.}{2020}]{redingetal20-1}
{Reding} J.~S.,  {Hermes} J.~J.,  {Vanderbosch} Z.,  {Dennihy} E.,  {Kaiser}
  B.~C.,  {Mace} C.~B.,  {Dunlap} B.~H.,   {Clemens} J.~C.,  2020, \mn@doi
  [\apj] {10.3847/1538-4357/ab8239}, \href
  {https://ui.adsabs.harvard.edu/abs/2020ApJ...894...19R} {894, 19}

\bibitem[\protect\citeauthoryear{{Ronco}, {Schreiber}, {Giuppone}, {Veras},
  {Cuadra}  \& {Guilera}}{{Ronco} et~al.}{2020}]{roncoetal20-1}
{Ronco} M.~P.,  {Schreiber} M.~R.,  {Giuppone} C.~A.,  {Veras} D.,  {Cuadra}
  J.,   {Guilera} O.~M.,  2020, \mn@doi [\apjl] {10.3847/2041-8213/aba35f},
  \href {https://ui.adsabs.harvard.edu/abs/2020ApJ...898L..23R} {898, L23}

\bibitem[\protect\citeauthoryear{{Schreiber}, {Belloni}, {G{\"a}nsicke},
  {Parsons}  \& {Zorotovic}}{{Schreiber} et~al.}{2021}]{schreiberetal21-1}
{Schreiber} M.~R.,  {Belloni} D.,  {G{\"a}nsicke} B.~T.,  {Parsons} S.~G.,
  {Zorotovic} M.,  2021, \mn@doi [Nature Astronomy, https://rdcu.be/cjFXN]
  {10.1038/s41550-021-01346-8}

\bibitem[\protect\citeauthoryear{{Smallwood}, {Martin}, {Livio}  \&
  {Lubow}}{{Smallwood} et~al.}{2018}]{smallwoodetal18-1}
{Smallwood} J.~L.,  {Martin} R.~G.,  {Livio} M.,   {Lubow} S.~H.,  2018,
  \mn@doi [\mnras] {10.1093/mnras/sty1819}, \href
  {https://ui.adsabs.harvard.edu/abs/2018MNRAS.480...57S} {480, 57}

\bibitem[\protect\citeauthoryear{{Stephan}, {Naoz}, {Gaudi}  \&
  {Salas}}{{Stephan} et~al.}{2020}]{alexanderetal20-1}
{Stephan} A.~P.,  {Naoz} S.,  {Gaudi} B.~S.,   {Salas} J.~M.,  2020, \mn@doi
  [\apj] {10.3847/1538-4357/ab5b00}, \href
  {https://ui.adsabs.harvard.edu/abs/2020ApJ...889...45S} {889, 45}

\bibitem[\protect\citeauthoryear{{Subasavage} et~al.,}{{Subasavage}
  et~al.}{2017}]{subasavageetal17-1}
{Subasavage} J.~P.,  et~al., 2017, \mn@doi [\aj] {10.3847/1538-3881/aa76e0},
  \href {https://ui.adsabs.harvard.edu/abs/2017AJ....154...32S} {154, 32}

\bibitem[\protect\citeauthoryear{{Swan}, {Farihi}, {Koester}, {Hollands},
  {Parsons}, {Cauley}, {Redfield}  \& {G{\"a}nsicke}}{{Swan}
  et~al.}{2019}]{swanetal19-1}
{Swan} A.,  {Farihi} J.,  {Koester} D.,  {Hollands} M.,  {Parsons} S.,
  {Cauley} P.~W.,  {Redfield} S.,   {G{\"a}nsicke} B.~T.,  2019, \mn@doi
  [\mnras] {10.1093/mnras/stz2337}, \href
  {https://ui.adsabs.harvard.edu/abs/2019MNRAS.490..202S} {490, 202}

\bibitem[\protect\citeauthoryear{{Tremblay}, {Cummings}, {Kalirai},
  {G{\"a}nsicke}, {Gentile-Fusillo}  \& {Raddi}}{{Tremblay}
  et~al.}{2016}]{tremblayetal16-1}
{Tremblay} P.~E.,  {Cummings} J.,  {Kalirai} J.~S.,  {G{\"a}nsicke} B.~T.,
  {Gentile-Fusillo} N.,   {Raddi} R.,  2016, \mn@doi [\mnras]
  {10.1093/mnras/stw1447}, \href
  {https://ui.adsabs.harvard.edu/abs/2016MNRAS.461.2100T} {461, 2100}

\bibitem[\protect\citeauthoryear{{Tremblay} et~al.,}{{Tremblay}
  et~al.}{2020}]{tremblayetal20-1}
{Tremblay} P.~E.,  et~al., 2020, \mn@doi [\mnras] {10.1093/mnras/staa1892},
  \href {https://ui.adsabs.harvard.edu/abs/2020MNRAS.497..130T} {497, 130}

\bibitem[\protect\citeauthoryear{{Van~Horn}}{{Van~Horn}}{1968}]{vanhorn68-1}
{van Horn} H.~M.,  1968, \mn@doi [ApJ] {10.1086/149432}, \href
  {https://ui.adsabs.harvard.edu/abs/1968ApJ...151..227V} {151, 227}

\bibitem[\protect\citeauthoryear{Vanderburg et~al.,}{Vanderburg
  et~al.}{2015}]{vanderburgetal15-1}
Vanderburg A.,  et~al., 2015, \mn@doi [Nature] {10.1038/nature15527}, 526,
  546–549

\bibitem[\protect\citeauthoryear{{Veras} \& {G{\"a}nsicke}}{{Veras} \&
  {G{\"a}nsicke}}{2015}]{veras+gaensicke15-1}
{Veras} D.,  {G{\"a}nsicke} B.~T.,  2015, \mn@doi [\mnras]
  {10.1093/mnras/stu2475}, \href
  {https://ui.adsabs.harvard.edu/abs/2015MNRAS.447.1049V} {447, 1049}

\bibitem[\protect\citeauthoryear{{Villaver} \& {Livio}}{{Villaver} \&
  {Livio}}{2009}]{villaver+livio09-1}
{Villaver} E.,  {Livio} M.,  2009, \mn@doi [\apjl]
  {10.1088/0004-637X/705/1/L81}, \href
  {https://ui.adsabs.harvard.edu/abs/2009ApJ...705L..81V} {705, L81}

\bibitem[\protect\citeauthoryear{{Walters} et~al.,}{{Walters}
  et~al.}{2021}]{waltersetal21-1}
{Walters} N.,  et~al., 2021, \mn@doi [\mnras] {10.1093/mnras/stab617}, \href
  {https://ui.adsabs.harvard.edu/abs/2021MNRAS.tmp..638W} {}

\bibitem[\protect\citeauthoryear{{Zuckerman} \& {Becklin}}{{Zuckerman} \&
  {Becklin}}{1987}]{zuckerman+becklin87-1}
{Zuckerman} B.,  {Becklin} E.~E.,  1987, \mn@doi [\nat] {10.1038/330138a0},
  \href {https://ui.adsabs.harvard.edu/abs/1987Natur.330..138Z} {330, 138}

\bibitem[\protect\citeauthoryear{{Zuckerman}, {Koester}, {Dufour}, {Melis},
  {Klein}  \& {Jura}}{{Zuckerman} et~al.}{2011}]{zuckermanetal11-1}
{Zuckerman} B.,  {Koester} D.,  {Dufour} P.,  {Melis} C.,  {Klein} B.,   {Jura}
  M.,  2011, \mn@doi [\apj] {10.1088/0004-637X/739/2/101}, \href
  {https://ui.adsabs.harvard.edu/abs/2011ApJ...739..101Z} {739, 101}

\makeatother
\end{thebibliography}

%\begin{thebibliography}{99}
%\end{thebibliography}

%%%%%%%%%%%%%%%%%%%%%%%%%%%%%%%%%%%%%%%%%%%%%%%%%%

%%%%%%%%%%%%%%%%% APPENDICES %%%%%%%%%%%%%%%%%%%%%

%\appendix



%%%%%%%%%%%%%%%%%%%%%%%%%%%%%%%%%%%%%%%%%%%%%%%%%%
\vspace{-0.25cm}

% Don't change these lines
\bsp	% typesetting comment
\label{lastpage}
\end{document}